\documentstyle[12pt]{article}
\newcommand{\nc}{\newcommand}
\nc{\renc}{\renewcommand}

\nc{\half}{{\textstyle{1\over2}}}
\nc{\etal}{\mbox{\it et al. }}
\nc{\ie}{{\it i.e.}}
\nc{\eg}{{\it e.g.}}

\renc{\thefootnote}{\arabic{footnote}}
\nc{\capt}[1]{{\bf Figure.} {\small\sl #1}}

\nc{\eqs}[2]{\mbox{Eqs.~(\ref{#1},\,\ref{#2})}}
\nc{\eq}[1]{\mbox{Eq.~(\ref{#1})}}

\nc{\figs}[2]{\mbox{Figs.~(\ref{#1},\,\ref{#2})}}
\nc{\fig}[1]{\mbox{Fig~.(\ref{#1})}}

\nc{\tag}[1]{\label{#1} \marginpar{{\footnotesize #1}}}
\nc{\mtag}[1]{\label{#1} \mbox{\marginpar{{\footnotesize #1}}}}
\renc{\baselinestretch}{1.5}
\newlength{\overeqskip}
\newlength{\undereqskip}
\nc{\be}[1]{\begin{equation} \mbox{$\label{#1}$}}
\nc{\bea}[1]{\begin{eqnarray} \mbox{$\label{#1}$}}
\nc{\Section}[2]{\section{#2}\label{#1}}
\nc{\Bibitem}[1]{\bibitem{#1}}
\nc{\Label}[1]{\label{#1}}

\nc{\eea}{\vspace{\undereqskip}\end{eqnarray}}
\nc{\ee}{\vspace{\undereqskip}\end{equation}}
\nc{\bdm}{\begin{displaymath}}
\nc{\edm}{\end{displaymath}}
\nc{\dpsty}{\displaystyle}
\nc{\bc}{\begin{center}}
\nc{\ec}{\end{center}}
\nc{\ba}{\begin{array}}
\nc{\ea}{\end{array}}
\nc{\bab}{\begin{abstract}}
\nc{\eab}{\end{abstract}}
\nc{\btab}{\begin{tabular}}
\nc{\etab}{\end{tabular}}
\nc{\bit}{\begin{itemize}}
\nc{\eit}{\end{itemize}}
\nc{\ben}{\begin{enumerate}}
\nc{\een}{\end{enumerate}}
\nc{\bfig}{\begin{figure}}
\nc{\efig}{\end{figure}}
\nc{\arreq}{&\!=\!&}
\nc{\arrmi}{&\!-\!&}
\nc{\arrpl}{&\!+\!&}
\nc{\arrap}{&\!\!\!\approx\!\!\!&}
\nc{\non}{\nonumber\\*}
\nc{\align}{\!\!\!\!\!\!\!\!&&}

\def\lsim{\; \raise0.3ex\hbox{$<$\kern-0.75em
      \raise-1.1ex\hbox{$\sim$}}\; }
\def\gsim{\; \raise0.3ex\hbox{$>$\kern-0.75em
      \raise-1.1ex\hbox{$\sim$}}\; }
\nc{\DOT}{\hspace{-0.08in}{\bf .}\hspace{0.1in}}
\nc{\Laada}{\hbox {$\sqcap$ \kern -1em $\sqcup$}}
\nc\loota{{\scriptstyle\sqcap\kern-0.55em\hbox{$\scriptstyle\sqcup$}}}
\nc\Loota{{\sqcap\kern-0.65em\hbox{$\sqcup$}}}
\nc\laada{\Loota}
\nc{\qed}{\hskip 3em \hbox{\BOX} \vskip 2ex}
\def\Re{{\rm Re}\hskip2pt}

\nc{\real}{{\rm I \! R}}
\nc{\Z}{{\sf Z \!\!\! Z}}
\nc{\complex}{{\rm C\!\!\! {\sf I}\,\,}}
\def\bigid{\leavevmode\hbox{\small1\kern-3.8pt\normalsize1}}
\def\id{\leavevmode\hbox{\small1\kern-3.3pt\normalsize1}}
\nc{\slask}{\!\!\!/}
\nc{\bis}{{\prime\prime}}
\nc{\pa}{\partial}
\nc{\na}{\nabla}
\nc{\ra}{\rangle}
\nc{\la}{\langle}
\nc{\goto}{\rightarrow}
\nc{\swap}{\leftrightarrow}

\nc{\EE}[1]{ \mbox{$\cdot10^{#1}$} }
\nc{\abs}[1]{\left|#1\right|}
\nc{\at}[2]{\left.#1\right|_{#2}}
\nc{\norm}[1]{\|#1\|}
\nc{\abscut}[2]{\Abs{#1}_{\scriptscriptstyle#2}}
\nc{\vek}[1]{{\rm\bf #1}}
\nc{\integral}[2]{\int\limits_{#1}^{#2}}
\nc{\inv}[1]{\frac{1}{#1}}
\nc{\dd}[2]{{{\partial #1}\over{\partial #2}}}
\nc{\ddd}[2]{{{{\partial}^2 #1}\over{\partial {#2}^2}}}
\nc{\dddd}[3]{{{{\partial}^2 #1}\over
	{\partial #2 \partial #3}}}
\nc{\dder}[2]{{{d #1}\over{d #2}}}
\nc{\ddder}[2]{{{d^2 #1}\over{d {#2}^2}}}
\nc{\dddder}[3]{{d^2 #1}\over
	{d #2 d #3}}
\nc{\dx}[1]{d\,^{#1}x}
\nc{\dy}[1]{d\,^{#1}y}
\nc{\dz}[1]{d\,^{#1}z}
\nc{\dl}[1]{\frac{d\,^{#1}l}{(2\pi)^{#1}}}
\nc{\dk}[1]{\frac{d\,^{#1}k}{(2\pi)^{#1}}}
\nc{\dq}[1]{\frac{d\,^{#1}q}{(2\pi)^{#1}}}

\nc{\cc}{\mbox{$c.c.$ }}
\nc{\hc}{\mbox{$h.c.$ }}
\nc{\cf}{cf.\ }
\nc{\erfc}{{\rm erfc}}
\nc{\Tr}{{\rm Tr\,}}
\nc{\tr}{{\rm tr\,}}
\nc{\pol}{{\rm pol}}
\nc{\sign}{{\rm sign}}
\nc{\bfT}{{\bf T }}

\def\GeV{{\rm\ GeV}}

\nc{\cA}{{\cal A}}
\nc{\cB}{{\cal B}}
\nc{\cD}{{\cal D}}
\nc{\cE}{{\cal E}}
\nc{\cG}{{\cal G}}
\nc{\cH}{{\cal H}}
\nc{\cL}{{\cal L}}
\nc{\cO}{{\cal O}}
\nc{\cT}{{\cal T}}
\nc{\cN}{{\cal N}}
\nc{\rvac}[1]{|{\cal O}#1\rangle}
\nc{\lvac}[1]{\langle{\cal O}#1|}
\nc{\rvacb}[1]{|{\cal O}_\beta #1\rangle}
\nc{\lvacb}[1]{\langle{\cal O}_\beta #1 |}
\nc{\bb}{\bar{\beta}}
\nc{\bt}{\tilde{\beta}}
\nc{\ctH}{\tilde{\cal H}}
\nc{\chH}{\hat{\cal H}}
\nc{\1}{\aa}
\nc{\2}{\"{a}}
\nc{\3}{\"{o}}
\nc{\4}{\AA}
\nc{\5}{\"{A}}
\nc{\6}{\"{O}}
\nc{\al}{\alpha}
\nc{\g}{\gamma}
\nc{\Del}{\Delta}
\nc{\e}{\epsilon}
\nc{\eps}{\epsilon}
\nc{\lam}{\lambda}
\nc{\om}{\omega}
\nc{\Om}{\Omega}
\nc{\ve}{\varepsilon}
\nc{\mn}{{\mu\nu}}
\nc{\k}{\kappa}
\nc{\vp}{\varphi}

\nc{\advp}[3]{{\it  Adv.\ in\ Phys.\ }{{\bf #1} {(#2)} {#3}}}
\nc{\annp}[3]{{\it  Ann.\ Phys.\ (N.Y.)\ }{{\bf #1} {(#2)} {#3}}}
\nc{\apl}[3]{{\it  Appl. Phys. Lett. }{{\bf #1} {(#2)} {#3}}}
\nc{\apj}[3]{{\it  Ap.\ J.\ }{{\bf #1} {(#2)} {#3}}}
\nc{\apjl}[3]{{\it  Ap.\ J.\ Lett.\ }{{\bf #1} {(#2)} {#3}}}
\nc{\app}[3]{{\it Astropart.\ Phys.\ }{{\bf #1} {(#2)} {#3}}}
\nc{\cmp}[3]{{\it  Comm.\ Math.\ Phys.\ }{{ \bf #1} {(#2)} {#3}}}
\nc{\cqg}[3]{{\it  Class.\ Quant.\ Grav.\ }{{\bf #1} {(#2)} {#3}}}
\nc{\epl}[3]{{\it  Europhys.\ Lett.\ }{{\bf #1} {(#2)} {#3}}}
\nc{\ijmp}[3]{{\it Int.\ J.\ Mod.\ Phys.\ }{{\bf #1} {(#2)} {#3}}}
\nc{\ijtp}[3]{{\it Int.\ J.\ Theor.\ Phys.\ }{{\bf #1} {(#2)} {#3}}}
\nc{\jmp}[3]{{\it  J.\ Math.\ Phys.\ }{{ \bf #1} {(#2)} {#3}}}
\nc{\jpa}[3]{{\it  J.\ Phys.\ A\ }{{\bf #1} {(#2)} {#3}}}
\nc{\jpc}[3]{{\it  J.\ Phys.\ C\ }{{\bf #1} {(#2)} {#3}}}
\nc{\jap}[3]{{\it J.\ Appl.\ Phys.\ }{{\bf #1} {(#2)} {#3}}}
\nc{\jpsj}[3]{{\it J.\ Phys.\ Soc.\ Japan\ }{{\bf #1} {(#2)} {#3}}}
\nc{\lmp}[3]{{\it Lett.\ Math.\ Phys.\ }{{\bf #1} {(#2)} {#3}}}
\nc{\mpl}[3]{{\it  Mod.\ Phys.\ Lett.\ }{{\bf #1} {(#2)} {#3}}}
\nc{\ncim}[3]{{\it  Nuov.\ Cim.\ }{{\bf #1} {(#2)} {#3}}}
\nc{\np}[3]{{\it  Nucl.\ Phys.\ }{{\bf #1} {(#2)} {#3}}}
\nc{\pr}[3]{{\it Phys.\ Rev.\ }{{\bf #1} {(#2)} {#3}}}
\nc{\pra}[3]{{\it  Phys.\ Rev.\ A\ }{{\bf #1} {(#2)} {#3}}}
\nc{\prb}[3]{{\it  Phys.\ Rev.\ B\ }{{{\bf #1} {(#2)} {#3}}}}
\nc{\prc}[3]{{\it  Phys.\ Rev.\ C\ }{{\bf #1} {(#2)} {#3}}}
\nc{\prd}[3]{{\it  Phys.\ Rev.\ D\ }{{\bf #1} {(#2)} {#3}}}
\nc{\prl}[3]{{\it Phys.\ Rev.\ Lett.\ }{{\bf #1} {(#2)} {#3}}}
\nc{\pl}[3]{{\it  Phys.\ Lett.\ }{{\bf #1} {(#2)} {#3}}}
\nc{\prep}[3]{{\it Phys\. Rep.\ }{{\bf #1} {(#2)} {#3}}}
\nc{\prsl}[3]{{\it Proc.\ R.\ Soc.\ London\ }{{\bf #1} {(#2)} {#3}}}
\nc{\ptp}[3]{{\it  Prog.\ Theor.\ Phys.\ }{{\bf #1} {(#2)} {#3}}}
\nc{\ptps}[3]{{\it  Prog\ Theor.\ Phys.\ suppl.\ }{{\bf #1} {(#2)} {#3}}}
\nc{\physa}[3]{{\it  Physica\ A\ }{{\bf #1} {(#2)} {#3}}}
\nc{\physb}[3]{{\it  Physica\ B\ }{{\bf #1} {(#2)} {#3}}}
\nc{\phys}[3]{{\it Physica\ }{{\bf #1} {(#2)} {#3}}}
\nc{\rmp}[3]{{\it  Rev.\ Mod.\ Phys.\ }{{\bf #1} {(#2)} {#3}}}
\nc{\rpp}[3]{{\it Rep.\ Prog.\ Phys.\ }{{\bf #1} {(#2)} {#3}}}
\nc{\sjnp}[3]{{\it Sov.\ J.\ Nucl.\ Phys.\ }{{\bf #1} {(#2)} {#3}}}
\nc{\spjetp}[3]{{\it Sov.\ Phys.\ JETP\ }{{\bf #1} {(#2)} {#3}}}
\nc{\yf}[3]{{\it Yad.\ Fiz.\ }{{\bf #1} {(#2)} {#3}}}
\nc{\zetp}[3]{{\it Zh.\ Eksp.\ Teor.\ Fiz.\  }{{\bf #1}  {(#2)} {#3}}}
\nc{\zp}[3]{{\it Z.\ Phys.\ }{{\bf #1} {(#2)} {#3}}}
\nc{\ibid}[3]{{\sl ibid.\ }{{\bf #1} {#2} {#3}}}
\nc{\rf}[1]{(\ref{#1})}
\nc{\nn}{\nonumber \\*}
\nc{\bfB}{\bf{B}}
\nc{\bfv}{\bf{v}}
\nc{\bfx}{\bf{x}}
\nc{\bfy}{\bf{y}}
\nc{\vx}{\vec{x}}
\nc{\vy}{\vec{y}}
\nc{\oB}{\overline{B}}
\nc{\oI}{\overline{I}}
\nc{\oR}{\overline{R}}
\nc{\rar}{\rightarrow}
\nc{\ti}{\times}
\nc{\slsh}{\hskip-5pt/}
\nc{\sm}{Standard~Model~}
\nc{\MP}{M_{\rm Pl}}
\nc{\tp}{t_{\rm Pl}}
\nc{\ave}{\bar{E}}

\nc{\eff}{{\rm eff}}
\nc{\kk}{\vek{k}}
\nc{\pp}{{\rm p}}
\nc{\ga}{g_{a\gamma}}
\nc{\vv}{\\}
\nc{\eee}{{\bf E}}
\nc{\bbb}{{\bf B}}
\nc{\qcd}{T_{\rm QCD}}
\nc{\G}{\rm \ G}
\def\vec#1{{\bf #1}}

\def\lae{\;^{<}_{\sim} \;} \def\gae{\; ^{>}_{\sim} \;} 
\def\tu{\tilde{\mu}}
\def\te{\tilde{\epsilon}}
\def\rd{\dot{R}}
\def\tee{\tilde{E}}

\begin{document}
{\title{\vskip-2truecm{\hfill {{\small \\
	}}\vskip 1truecm}
{\bf Can Cosmic Ray Catalysed Vacuum Decay 
Dominate Over Tunnelling?}}


{\author{
{\sc  Kari Enqvist$^{1}$ and John McDonald$^{2}$}\\
{\sl\small Department of Physics, P.O. Box 9,
FIN-00014 University of Helsinki,
Finland}
}
\maketitle
\vspace{1cm}
\begin{abstract}
\noindent
	We consider the question of whether 
cosmic ray catalysed false vacuum decay 
can be phenomenologically more important than 
spontaneous decay via quantum tunnelling. We 
extend the zero bubble wall width Landau-WKB 
analysis of catalysed false vacuum decay 
to include the leading order effects of finite wall width and derive
an expression for the thin-wall bubble action.  Using this we 
calculate the exponential suppression factor for the 
catalysed decay rate at the critical bubble energy, corresponding  
to the largest probability of catalysed decay.
We show that, in general, cosmic ray catalysed 
decay is likely to be more important than
spontaneous decay for sufficiently thin-walled bubbles (wall thickness less than about 
$30 \%$ of the initial bubble radius), but that spontaneous 
decay will dominate for the case of thick-walled bubbles.
Since any perturbative model with a cosmologically 
significant false vacuum decay rate will 
almost certainly produce thick-walled bubbles, 
we can conclude 
that cosmic ray catalysed false vacuum decay will 
never dominate over tunnelling in imposing 
phenomenological constraints on perturbative particle physics models. 
\end{abstract}
\vfil
\footnoterule
{\small $^1$enqvist@rock.helsinki.fi};
{\small $^2$mcdonald@phcu.helsinki.fi}

\thispagestyle{empty}
\newpage
\setcounter{page}{1}

\section{Introduction}

	   False vacuum decay \cite{cea} is known to play an important role in 
constraining the parameter space of many models of particle physics. 
Examples include the Standard Model itself \cite{smd}, which can have a 
metastable vacuum state for a range of the parameters which determine 
the Higgs potential, and the Minimal Supersymmetric extension of the
 Standard Model (MSSM) \cite{ssd,ebd}, 
which can have charge and colour breaking
 minima for a range of its parameters. In particular, the question of the 
stability of the electroweak vacuum with respect to decay to colour breaking 
minima is of particular significance for the possibility of 
successful 
electroweak baryogenesis within the framework of the MSSM \cite{ebd}. 

	 Usually it is assumed that the most important false vacuum 
decay mode is spontaneous decay via quantum tunnelling to the true 
vacuum state \cite{cea}. However, it has been recognized that false vacuum 
decay could also be catalysed, in principle, by collisions of cosmic rays
 which produce a virtual scalar field with sufficient energy to 
nucleate a super-critical bubble of false vacuum \cite{eea,v2}.
 Early attempts to estimate the rate 
of such catalysed events indicated that such processes might
 be more important than spontaneous decay for a range of the 
parameter space \cite{eea}. However, these discussions were based upon 
somewhat uncertain physical reasoning and assumptions, making it 
difficult to assess their validity. Later, the problem was approached
both numerically and analytically. The most recent numerical analysis, based 
on a lattice calculation of the decay rate, 
indicates that the exponential factor in catalysed decay is about 
0.8 times that in spontaneous decay \cite{tea}. 
This calculation was done in the context of a 
$ - \lambda \phi^{4}$ theory and so 
corresponds to the extreme case of a thick-walled bubble. 

		  In the present paper we will be concerned with the
 analytical treatment of catalysed false vacuum 
decay proposed by Voloshin \cite{v1}, 
based on the Landau-WKB method of calculating transitions between 
strongly different states in quantum mechanics \cite{ll}.
 In this approach one writes 
a Lagrangian for a bubble of true vacuum as a function of its radius $R$, 
$ L(R, \dot{R})$, and then quantizes this action and computes the
 amplitude for the creation of a bubble of radius $R$ from the initial false 
vacuum state by using the WKB method as developed by Landau.
 Our main goals in this paper will be 
to extend the Voloshin calculation, which was based on 
bubbles of vanishing wall thickness, to the 
case where the effects of bubble wall width are included and then to 
apply our results to the question of whether 
cosmic-ray catalysed false vacuum decay 
can be more important that spontaneous decay in imposing phenomenological 
constraints on particle physics models. Compared with the earlier attempts to address
this question, we believe that our approach provides a much more firm and 
quantitative answer.

		 The paper is organized as follows. In section 2 we
 review the application of the Landau-WKB method to  
false vacuum bubble formation 
in the presence of a virtual scalar field for the case of 
bubbles of zero wall width.
 In section 3 we derive an action for thin-walled 
bubbles including the leading order corrections 
due to finite wall width. In section 4 
we apply the Landau-WKB method
to this action and calculate the effect of finite wall 
width on the bubble formation
probability. In section 5 we discuss cosmic ray catalysed 
and spontaneous false 
vacuum decay and the phenomenological implications of our results. 
In section 6 we present and discuss our conclusions. In the Appendix 
we review the original Landau discussion of the application of the WKB 
approximation to the calculation 
of matrix elements of strongly different states in quantum mechanics.
\section{Virtual Scalar Field Catalysis of 
False Vacuum Decay in the Zero-Width Limit}

In this section we set up our notation and
review the application of the Landau-WKB 
method to the 
catalysis of false vacuum decay in the limit of vanishing bubble wall 
thickness \cite{v1}. We will consider a real scalar field with a
potential $U(\phi)$ of the form 
\be{pot}
U(\phi) = \frac{\lambda}{4} (\phi^2 - v^2)^2  - M^3 \phi  
\equiv U_{o}(\phi) - M^3 \phi  ~,
\ee
which has two minima with an energy difference $\epsilon = 2vM^3$. 
For further use we define
\be{tildes}
\te={4\pi\epsilon\over 3}~,~~\tu=\frac{8\pi}{3}\sqrt{2\lambda}v^3~.
\ee
It can be shown \cite{v2} that the classical  Lagrangian
for a zero-width bubble wall reads 
\be{lagr}
L(\dot{R}, R) = -\tu \sqrt{1-\rd^2} \; R^2 + \te R^3 ~,    
\ee
where $R=R(t)$ is the radius of the bubble. In the next section
we derive and generalize this result for non-zero bubble width.
The classical Lagrangian \eq{lagr} has one degree of freedom, which
may be canonically quantized as for the case of ordinary 
quantum mechanics. In particular, the WKB method of Landau for the
 calculation of the transition matrix element between two
 "strongly different" states \cite{ll} may be applied to \eq{lagr}.
 (We review the quantum mechanical Landau-WKB method in the Appendix). 

		  Landau showed that, up to a relatively 
unimportant prefactor, the transition matrix 
element of an operator $f$ between two states 
$ \psi_{1}$ and $ \psi_{2}$ ($ E_{2} > E_{1}$) is given by 
\be{f12}
f_{12} \sim \exp\left[ Re \left( \int_{a_{1}}^{x^{*}} 
\sqrt{2 m (U-E_{1})}dx^{´}   -
 \int_{a_{2}}^{x^{*}} \sqrt{2 m (U-E_{2})}dx^{´} \right) \right] ~,
\ee
where $ a_{1}$ and $ a_{2}$ are the classical 
turning points of the motion 
of the particles in the presence of a potential barrier
$ U(x)$ and $ x^{*}$ is the
"transition point", corresponding to a point in the complex plane at which 
the derivative of the exponent with respect to $ x^{*}$ is zero. 
This expression can be 
directly generalized to the matrix
 element of any operator between strongly different 
initial and final states \cite{v1}, 
\be{melement} 
\langle Y(E_{2})| f |X(E_{1})\rangle  \sim \exp\left[-I_{L}\right] 
\equiv \exp\left[ \Re \left(i \int_{q_{X}}^{q^{*}} 
p(q ; E_{1}) dq + i \int_{q^{*}}^{q_{Y}} p(q ; E_{2}) dq \right) \right]~, 
\ee
where $p$ are the canonical momenta,  $ q_{X}$  and $ q_{Y}$ 
 are the classical turning points and $ q^{*}$ is the transition point. 
In the following we will refer to $ I_{L}$ as the Landau integral. 

	     For the case of the bubble 
 Lagrangian of \eq{lagr}, the canonical momentum $ p_{R}$
 is given by 
\be{canp} 
p_{R} \equiv \frac{\partial L}{\partial \rd} 
= \frac{\tu R^{2} \rd}{\sqrt{1-\rd^2}}~,
\ee
and the Hamiltonian $ H = p_{R} \rd - L$ is given by 
\be{hamilton}
H = \frac{\tu R^{2}}{\sqrt{1-\rd^2}} - \te R^3~.
\ee
In order to evaluate the Landau integral, we must solve the Hamiltonian 
\eq{hamilton} in order to find  $ \rd$ as a function of $R$ and $E$ 
and so obtain $ p_{R}$ as a function of 
$R$ and $E$. This step turns out to be particularly easy 
for the zero-width case. We obtain
\be{zeowidth}
\rd^{2} = \frac{(\tee^2 - \tu^2 R^{4})}{\tee^2}
\ee
and
\be{zerowpr}
p_{R} = (\tee^2 - \tu^2 R^4)^{1/2}~,
\ee
where $ \tee = E + \te R^3$. 

The amplitude $<B(E)|0>_{virtual}$ for the transition  
from the false vacuum state 
to a bubble of energy $E$ in the presence of a  
virtual scalar field
is related to the matrix element of the  
scalar field $ <B(E)| \phi(x) |0>$ integrated over $x$. 
This can be seen by noting 
that a virtual scalar  
field is created by including a source term $J \phi$ in the Lagrangian. 
Thus, representing the amplitude by a path integral, 
we see that 
switching on the source $J$ will produce a change in the 
amplitude according to
\be{amp} \frac{\delta <B(E)|0>}{\delta J} = 
\frac{\delta}{\delta J} \int [d \phi] e^{-\int d^{4}x L + J\phi}
= <B(E)|\phi(x)|0> .~\ee 
In the effective theory of thin wall bubbles of radius $R$ 
the non-trivial part of the 
field operator integrated over all space becomes
\be{fop} \int d^{3}x (\phi(x)+v) = \frac{8 \pi v}{3}R^{3}      .~\ee
Thus the calculation of the matrix element of the scalar field operator 
reduces to the calculation of the matrix element of $R^3$ in the  
effective theory.
Thus in terms of the dimensionless variables $x,~ e$ and $ \xi$, defined by 
\be{dimless}
R = x \frac{\tu}{\te}~,~~E = e \frac{\tu^3}{\te^2}~,~~ \xi = 
\frac{\te^3}{\tu^4}~, 
\ee
the exponential suppression of the 
amplitude for the creation of a  
super-critical 
bubble of energy $E$ from the vacuum state (corresponding to a bubble
of zero energy and radius) in the presence of a virtual 
scalar field is given by \cite{v1} 
\be{ebubble}
\langle B(E)| R^{3}| 0\rangle \sim \exp\left( - 
\frac{1}{\xi} \Re \left[ \int_{0}^{x^{*}}
 \sqrt{x^4-x^6} dx + \int_{x^{*}}^{x (E)} 
\sqrt{x^4-(x^3+e)^2} dx \right]\right)~.
\ee
The turning point $ x(E)$, corresponding to the radius 
of a bubble of energy $E$, is obtained by setting 
$ \rd = 0$ in the Hamiltonian \eq{hamilton}, which gives
\be{xe2}
x(E)^2 - x(E)^3 = e   ~.
\ee
We note that,  
whereas in the original Landau derivation 
the positive square root should be taken in the canonical momentum 
(see the Appendix), for 
the case of super-critical bubble formation from a sub-critical bubble 
initial state the 
directions of the incoming momenta of the initial and final bubble states are 
opposite, implying that the negative square root should be taken in 
the second integral. This is also necessary in order to obtain a 
non-trivial complex transition point. 
The transition point is then given by the solution of
\be{transitionp} 
\sqrt{x^4-x^6} = \sqrt{x^4 - (x^3 + e)^2}  ~.
\ee
Thus in the zero width case 
the transition point is given by $ x^{*}(e) = (-{e}/{2})^{1/3}$. 
Integrating up to the root in the upper half-plane, Voloshin 
showed that the smallest exponential suppression factor,
 corresponding to the largest 
probability of catalysed vacuum decay, is obtained when the energy 
corresponds to the maximum energy for which a sub-critical bubble can 
exist i.e. the energy of the critical "sphaleron" 
bubble configuration at the top of the 
energy barrier seperating the sub-critical and super-critical 
bubble regions 
\cite{v1}. 
From $ H(\rd = 0)$ we see that 
this corresponds to a bubble of radius $ x_{c} = 2/3$ and energy 
$ e_{c} = 4/27 \approx 0.15$. The exponential 
suppression factor is then given by \cite{v1}
\be{suppress}
|\langle B(E_{c}))| R^3 | 0\rangle |^2 \sim \exp\left(-F(E{c})\right)    ~, 
\ee
where 
\be{FEc}
\frac{F(E_{c})}{F(0)} \approx 0.16     ~
\ee
and $F(0)$ corresponds to the bounce action for vacuum tunnelling \cite{cea}.
For energies greater than the critical energy the
 additional energy will be dissipated via perturbative particle production (which is 
favoured over formation of a higher energy bubble), 
such that no significant increase in the vacuum decay probability is obtained
for $ E > E_{c}$ \cite{v1}. 

     In the above calculation the exponential 
suppresion factor is, in fact, independent of the scalar field operator 
in the matrix element. Provided that the operator does not 
introduce any significant exponential factor, the details of 
the operator will only influence the 
pre-factor, which will play only a small role in 
constraining the parameters of a given
model. 

		  Physically, the importance of
 the virtual scalar field is that it provides a
 source of energy, making possible the transition to
a critical bubble of true vacuum.
In order to avoid further large suppression factors 
we should require that the virtual scalar field is able to efficiently 
supply energy to the region where the bubble is forming. 
This will be true if the region
 in which the energy of the virtual scalar field 
is concentrated is smaller than the size of the critical bubble. 
\section{Width-Corrected Thin Wall Action} 

		 In models 
of phenomenological interest the region of parameter space where 
false vacuum decay can be cosmologically significant is close to the thick wall limit
\cite{smd,ssd,ebd}. 
Therefore if we wish to apply the Landau-WKB method to cases of 
interest to phenomonology then
 we must extend the above calculation in order to take into 
account the effects of finite wall width. To do 
this we first need to derive a thin wall 
Lagrangian $ L(R, \rd)$ which includes the leading order corrections 
due to finite wall width. 

	Following Voloshin \cite{v1}, we will consider 
a real scalar field with a 
potential given by \eq{pot}. 
In the limit of small enough $M$ and bubble 
radius much larger than the wall width, 
the thin-wall solution of the $ \phi$ equation of 
motion may be written as \cite{cea} 
\be{thinwall} 
\phi (\rho) = - v \; {\rm Tanh}\left[\frac{1}{w} (\rho-k)\right]     ~,
\ee
where $ \rho^2 = r^2 - t^2$, $w$ is the half width of the
 wall and $k$ parameterizes the initial radius of the bubble. 
In order 
to calculate the width corrections to $ L(R,\rd)$ in the thin-wall limit 
(where $R$ is now defined to be the radius at the point in the bubble
 wall at which $ \phi = 0$) we 
substitute the solution \eq{thinwall} into the scalar field action.
This is given by
\be{subst1}
I = -4 \pi \int dt \; dr\;  r^2 
\left[\frac{1}{2} \left(\frac{\partial \phi}{\partial \rho}\right)^2
 + U(\rho) \right]     ~.
\ee
We write this as $ I = I_{g} + I_{\epsilon}$, where 
\be{subst2} 
I_{g} =  -4 \pi \int  dt \; dr\; r^2 
\left[\frac{1}{2} \left(\frac{\partial \phi}{\partial \rho}
\right)^2 
+ U_{o}(\rho) \right]  
\ee
and  
\be{subst3} 
I_{\epsilon} =  -4 \pi \int dt \; dr \; r^2 
\left(-M^3 \phi\right)  ~. 
\ee
 On substituting the thin wall solution into these we obtain 
\bea{subst4} 
I_{g} &=&-2 \pi \lambda v^4 \int dt \; 
d \rho \; \rho (\rho^2 + t^2)^{1/2} {\rm Sech}^{4}
\left[\frac{1}{w}(\rho-k)\right]  ~, \nn
I_{\epsilon} &=&-4 \pi M^3 v \int dt \; dr 
\; r^2 {\rm Tanh}\left[ \frac{1}{w}(\rho - k)\right]  ~. 
 \eea
   Evaluating these integrals then gives us a 
Lagrangian as a function of $k$ and $t,~ L(k,t)$.        
Since for the thin-wall solution \eq{thinwall} 
we know that the radius is given by 
\be{radius}
R(t) = (k^2 +  t^2)^{1/2}     ~,
\ee
we see that k and t may be written in terms of R and $\rd$ as
\be{koo}
k = R \sqrt{1-\rd^2}   ~
\ee
and
\be{tee}
t = R \rd ~.
\ee
Substituting these into $L(k,t)$  then gives us the desired Lagrangian 
$ L(R,\rd)$ for which \eq{radius} corresponds 
to a minimum action solution. 

	     In order to obtain an analytic expression 
for the action from \eq{subst4}
 we must fit the 
$ {\rm Sech}^{4}(x)$ and ${\rm Tanh}(x)$ functions 
with approximate polynomial expansions.
In practice we will use 
the expansion Sh4$(x)$ for $ {\rm Sech^{4}}(x)$ and 
Th$(x)$ for Tanh(x), where Sh4$(x)$ is defined by
\be{sh4} 
{\rm Sh4}(x) = \cases{{
\frac{984}{1000} - \frac{1601}{1000}x^{2} + 
 \frac{970}{1000}x^{4} - \frac{1987}{10000}x^{6}~~~ (|x| < 1.5)}\cr
{0~~~ (|x| \ge 1.5)~,}\cr}
\ee
and Th$(x)$ is defined by
\be{th} 
{\rm Th}(x) =\cases{{
\frac{9937}{10000}x - \frac{28597}{100000}x^{3}
 +  \frac{62238}{1000000}x^{5} - \frac{571}{100000}x^{7}~~~(|x| < 2)}\cr
	{-1~~~(x\le -2)}\cr
	{1~~~(x\ge 2)}~.\cr}
\ee
In order to obtain the leading order width corrections, 
we expand the integrals in terms of $ \frac{w}{R}$.
Such an expansion is possible for $ \left|\frac{w}{R}\right|$
small compared with 4, which is also necessary for the thin wall
solution for $ \phi$  \eq{thinwall} to be valid.

		Integrating \eq{subst4}, substituting \eqs{koo}{tee}
 for $k$ and $t$ and expanding in powers of $w$ 
then gives the width corrected thin wall Lagrangian,
 \be{wallact}
L(R,\rd)  = - \tu \sqrt{1-\rd^{2}} \left[R^{2} + a w^{2} 
\left(1 +
 \frac{\rd^{2}}{2}\right)\right]+ 
\te R^3 + b \te R w^{2} \left(1 - \frac{\rd^2}{2}
\right) + {\cal O}(w^4) ~,
\ee
 which is valid for $ |R| \gg 4w$ and where $a=0.29$ and $b=2.21$.
 This reduces to the zero-width Lagrangian \eq{lagr} in the $ w \rightarrow 0$ limit.
\section{Landau-WKB Calculation of the Catalysed 
Decay Rate for Non-Zero Width}

       In order to calculate the rate of catalysed vacuum decay we follow 
the same steps as for the zero-width calculation. The 
canonical momentum and Hamiltonian are now given by 
\be{canmom}
p_{R} = \frac{\tu \rd}{\sqrt{1-\rd^{2}}} \left[R^{2} +
 c w^{2} \rd^{2}\right] -b \te R \rd^{2}   ~,
\ee
where $c=0.44$, and 
\be{energy}
H + \te R^{3} = \frac{\tu}{\sqrt{1-\rd^{2}}} \left[R^{2} +
 a w^{2}(1 - \frac{\rd^2}{2}+ \dot{R}^{4}) \right] 
-b \te R  (1 + \frac{\rd^2}{2}) w^{2}   
~.
\ee
The parameters of the potential \eq{pot} are related to the
parameters appearing in the Hamiltonian \eq{energy} by
\be{relat}
v=\left(\frac{3}{16\pi} w\tu\right)^{1/2}~;~~ 
\lambda=\left(\frac{32\pi}{3}\frac{1}{\tu w^3}\right)~;~~  
M^3=\left(\frac{3}{4\pi}\right)^{1/2}{\te\over (w\tu)^{1/2}}~. 
\ee

We then solve the Hamiltonian equation in order 
to obtain $ \rd$ as a function of $R$ and $E$. 
In general we must solve
\be{mustsolve}
\left(\tee + b \te R \rd 
\left(1 + \frac{\rd^2}{2}\right) w^{2} \right)^{2}(1-\rd^{2}) 
= \tu^2  \left[R^{2} + a w^{2}(1 - 
\frac{\rd^2}{2}+ \dot{R}^{4}) \right]^{2}   
    ~.
\ee
This will give a 4th order polynomial in $ y = \rd^{2}$. To simplify
 this and so obtain an analytical expression, 
we will restrict attention to the 
case where the $ w^{4}$ terms on either side can be neglected.
In this case \eq{mustsolve} reduces to
\be{reduces}
f_{2} y^{2} + f_{1} y + f_{0} = 0   ~
\ee
where
\bea{ff}
f_{2} &=& (b \te \tee R + 2a\tu^{2} R^{2}) w^{2}~, \nn
f_{1} &=& (\tee^{2} + b \te \tee R w^{2}
- a \tu^{2} R^{2} w^{2} )  ~,\nn
 f_{0} &=& - \tee^{2} + \tu^{2} R^{4} - 2b \te \tee R 
w^{2}+2a \tu^{2} R^{2} w^{2}   ~.
\eea
Thus to this order one finds that
\be{rdee}
\rd^{2} = \frac{1}{2 f_{2}} \left[ - f_{1} + 
\sqrt{f_{1}^{2} - 4 f_{2} f_{0}} \right]~,  
\ee
where 
\bea{fonesquared}  \nonumber
f_{1}^{2} - 4 f_{2} f_{0} &=&\tee^{4} + 
(13.26 \te \tee^{3} R +
11.74 \tee^{2}\tu^{2} R^{2}-8.84 \te \tee 
\tu^{2} R^{5} -2.32 \tu^{4} R^{6}) w ^{2} \nn
&+& (43.96 \te^{2} \tee^{2} R^{2} + 3.84 \te \tee \tu^{2} R^{3}
-1.26 \tu^{4} R^{4}) w^{4}   ~. 
\eea
For sufficiently small width $ \rd^{2}$ 
can be expanded in terms of $ w^{2}$
\be{rdotexp}
 \rd^{2} = \rd_0^{2} + w^{2} \dot{R}_{1}^{2}    ~,
\ee
where $ \dot{R}_0^{2}$ is the zero width expression
given in \eq{zeowidth}. 
In addition, so long as $ \dot{R}_0^{2}$ is large compared with 
$ w^{2} \dot{R}_{1}^{2}$, we can expand $ \dot{R}_0$  
in powers of $ w^{2}$, 
\be{approdotR}
\dot{R} \approx \rd_0 \left(1 + \frac{w^{2}
 \rd_{1}^{2}}{2 \rd_0^{2}}\right)    ~.
\ee
We will also assume that we can expand $ (1-\rd^2)^{1/2}$
 in powers of $ w^{2}$.
In this case we can give an analytic expression for the shift in 
the transition point $ x^{*}$ relative to 
that obtained in the zero-width limit
$ x_0^{*} = \left(-{e}/{2}\right)^{1/3}$. 
In terms of the rescaled variables $x,~ e$ and 
$ d =  w{\te}/{\tu}$, 
 the canonical momentum is given by
\be{pxe} 
p(x,e) = p_0(x,e) \left(1 + 
d^{2} g(x,e)\right) - a x \rd_0 d^{2}   ~,
\ee
where 
\be{deff} 
g(x,e) = \frac{1}{2} \; \frac{\rd_{1}^{2}}{\rd_0^{2}
 (1 - \rd_0^{2}) }+ \frac{c \rd_0^{2}}{x^{2}}   ~.
\ee
The transition point $ x^{*}(e)$ 
is then given by the solution of 
$ p(x,0) = p(x,e)$. We define $ x_{1}(e)$ by
\be{transpn}
x^{*}(e) = x_0^{*}(e) + d^{2} x_{1}(e)~.
\ee
 With the expansion $ p(x^{*},e) = p_0(x^{*},e) 
+ d^{2} x_{1} p_0^{'}(x^{*},e)$, where
\be{ppilkku}\nonumber
p_0^{'}(x,e) = {(6 \te x^{2} 
- 4 x^{3})\over (2 \sqrt{\te^{2}-x^{4}})} ~,
\ee
we obtain for the shift in the transition point 
\be{shift} 
x_{1}(e) = -\frac{1.105}{\left(
-{e}/{2}\right)^{1/3}} - \frac{0.737}{e}     ~.
\ee
The turning point $x(E)$ is given by $H(\rd=0) = E$, so that
\eq{energy} implies that
\be{turning}
x(E)^3-x(E)^2+e+bx(E)d^2-ad^2=0~,
\ee
which can be solved numerically with $a$ and $b$ as in \eq{wallact}.

To obtain the exponential suppression factor for catalysed decay,
we evaluate the Landau integrals at the critical energy $e_c$
so that in the scaled variables given by \eq{dimless}
\be{ours}
{\xi}F(e_c,d)= \Re \left[ \int_{0}^{x^{*}(e_c)}
p(x,0)dx+\int_{x^{*}(e_c)}^{x(E_c)}p(x,e_c)dx\right]
\ee
with $p(x,e_{c})$ in general given by \eq{pxe}. 
       In the evaluation of the $e$=0 part of the  
Landau integral \eq{ours}
there are two distinct regions. For 
$ |x| \gae 1.15 d^{1/3}$ we can expand $ \rd^{2}$ and so $ p(x,0)$ 
in powers of $ d^2$ as in \eq{pxe}. For $ 4 d \lae |x| \lae 1.15 d^{1/3}$
 this is no longer possible and we find from \eq{rdee} that 
$ \rd^2$ becomes
\be{rdappro}
\rd^2 \approx i \; \frac{1.3 x}{d}   ~, 
\ee
where $ \sqrt{-1}$ is taken to equal $ +i$ 
in accordance with our definition of the 
phases in the complex plane. Thus the 
canonical momentum in this case is given by 
\be{px0} 
p(x,0)  = \frac{\tu^3}{\te^2} 
\left[ (-i x^2 + 0.57 xd) - 2.52 i^{1/2} x^{3/2} d^{3/2}
\right]    ~.
\ee

We have integrated \eq{ours} numerically, using the above expressions for 
$p(x,e)$. The thin-wall approximation breaks down
when $|x|$ is smaller than $d$, and therefore we  have used a
small $x$ cut-off at $ |x| = 4d$ in the $d$-dependent part
of $p(x,e)$. This does not affect the final results appreciably.
Some care must be taken in performing the
complex integrals, in particular when the integration path
is crossing branch cuts. Our result for the exponential suppression factor 
$ F(E_{c},d)$, together with $ F(0,d)$ which corresponds to the bounce action,
 is shown in Figure 1. At $d=0$ both curves agree with 
previous results \cite{v1}.
	     
The quantity of particular interest to us here is
the ratio of the exponential suppression factor 
at zero energy to that at the critical energy as a function of the 
wall width $d$. This measures the importance of catalysed decay, and
we parameterize this by $ \kappa (d)$, 
where 
\be{defkappa}  
F(E_{c},d) = 0.16 \kappa(d) F(0,d)     ~.
\ee
In the limit of zero width $ \kappa(0) =1$ \cite{v1}.

 \begin{figure}
\leavevmode
\centering
\vspace*{70mm} 
\includegraphics{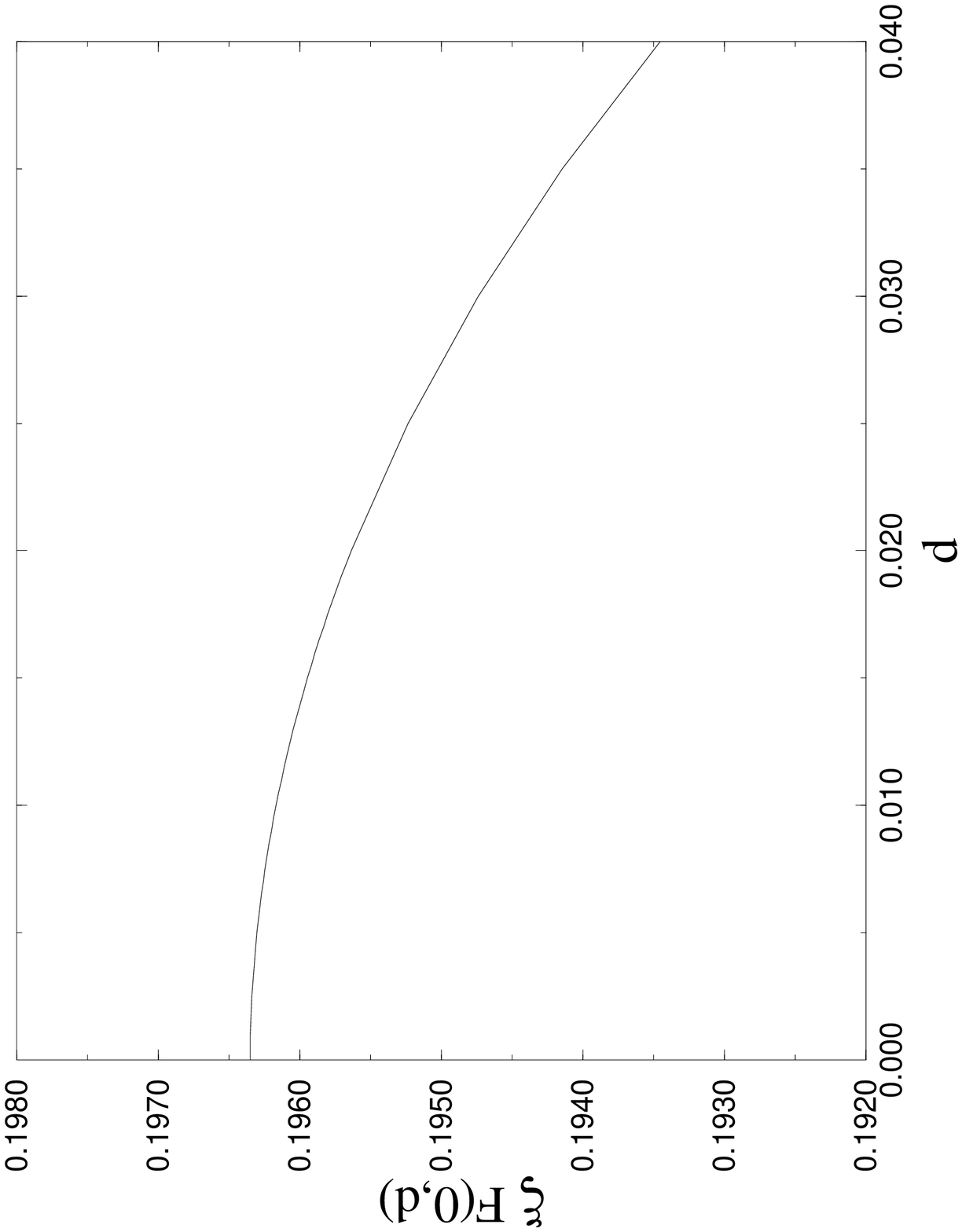}
\includegraphics{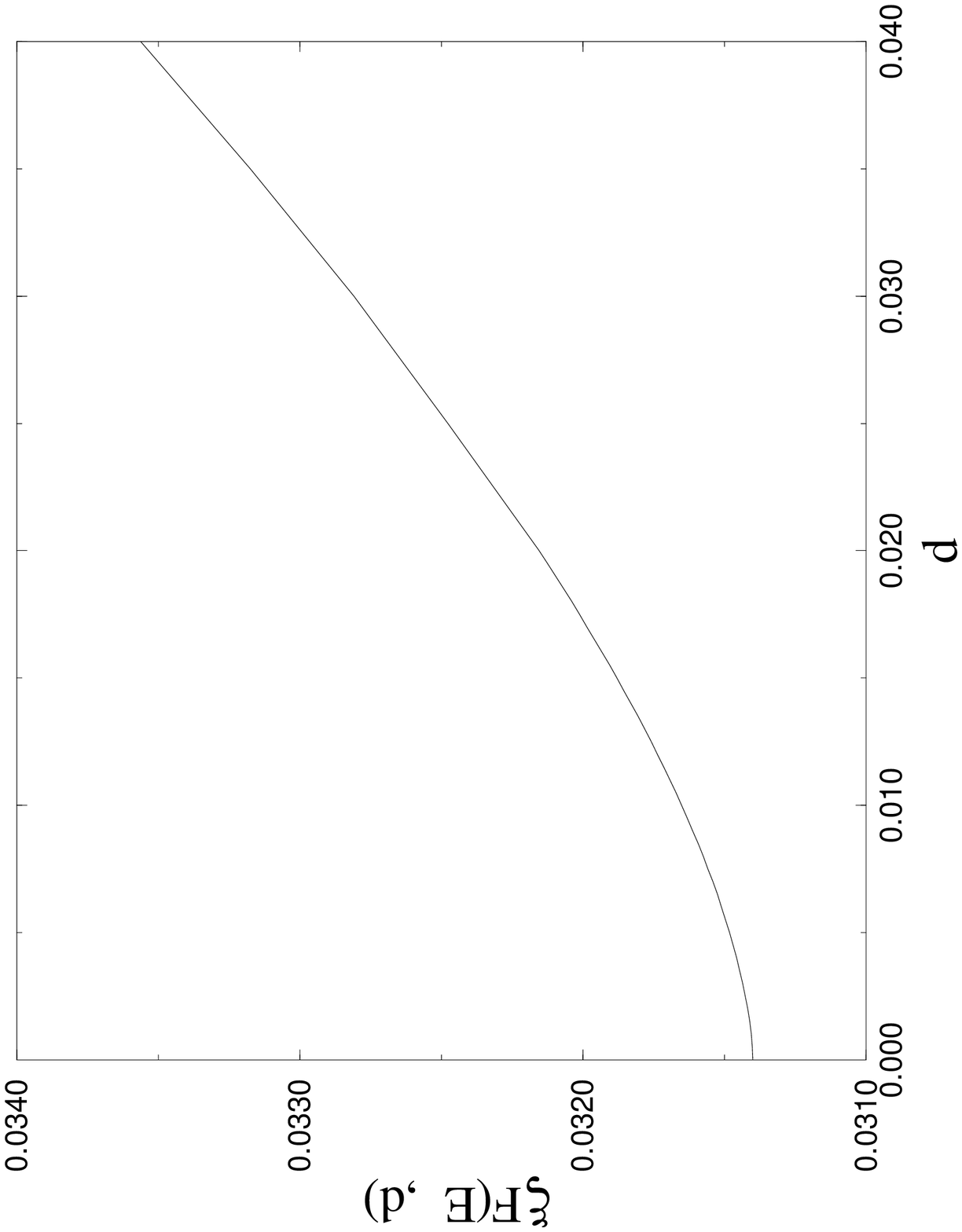}   
\caption{The scaled bounce action $\xi F(0,d)$ and the scaled action at the critical
energy $\xi F(E_c,d)$ as functions of the scaled width $d$.}
\label{kuva1}       
\end{figure} 

We display $\kappa(d)$ in Figure 2, and one can
see that $ \kappa(d)$ and the exponential suppression factor for catalysed decay 
increase as
 $d$ increases, whilst the 
bounce action decreases. 
The increase in $ \kappa(d)$ with increasing $d$ is  
in agreement with trend suggested by 
the numerical results for the thick wall bubble obtained by 
Kuznetsov and Tinyakov \cite{tea}, which indicate a value 
$ \kappa(d) \approx 5$ for the case of the thick-walled
$ -\lambda \phi^4$ theory 
they considered. We have tested our results by 
comparing the $d$ dependence of $ 
F(0,d)$ with that of the equivalent Euclidean bounce 
action calculated directly from the scalar field 
solution corresponding to a given 
value of $d$.
We find that the Landau-WKB result at $ E=0$ 
does indeed reproduce the 
bounce action for all $d$, up to the limits of our approximations. 
 \begin{figure}
\leavevmode
\centering
\vspace*{65mm} 
\includegraphics{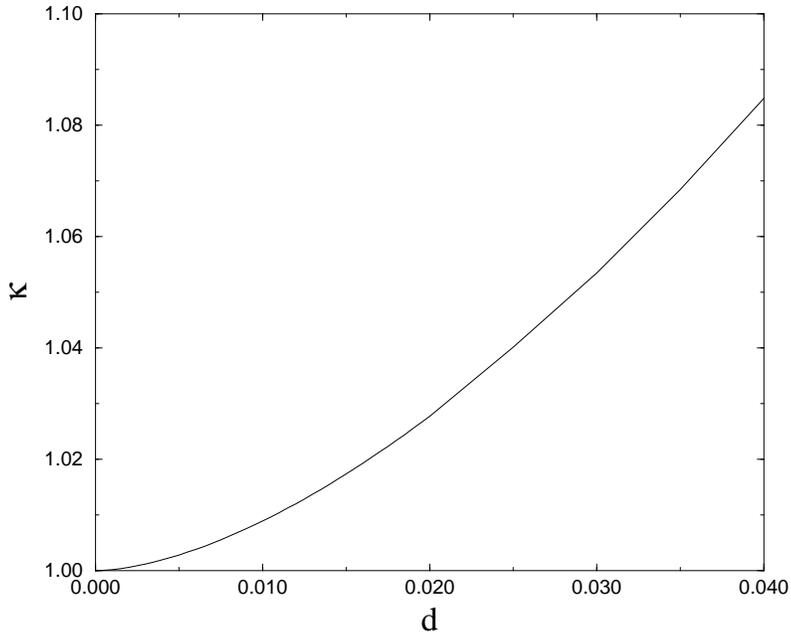}
\caption{The parameter $\kappa$ as a function of $d$.}
\label{kuva2}       
\end{figure} 

       In obtaining these results we have made a number of approximations, which 
impose limits on the range of wall widths we are able to discuss. We find that 
the tightest constraint comes
 from the requirement that the expansion of $\rd^2$ in powers 
of $w$, \eq{approdotR}, is valid at the transition point,
 which requires, for the $e = 0$ part of the Landau integral, 
that $ 2.32 d^2 \ll |x^{*}(e_{c})|^{6}$ for $|x^{*}(e_{c})|
 \approx |x^{*}_{0}(e_{c})| = 0.43 $. 
This requires that $d$ is small compared with 0.05. 
Although this is a tight constraint
on the range of widths for which our calculation is valid, 
physically we do not expect
any change in the d dependence of $\kappa (d)$ at $d \approx 0.05$. 
We should be 
able to extrapolate the $d$ dependence of $ \kappa (d)$
we obtain at small values of $d$, for which our calculation is valid,
up to larger values of $d$ so long as the wall width is small 
compared with the radius of the bubble, which provides the only other
length scale in the problem. 
Fitting 
$\kappa(d)$ with a polynomial up to $d^4$, using values of $d$ up to 0.01, we obtain 
\be{kdfit} \kappa(d) \approx 0.999 + 0.125 d + 110.3 d^2  +  (-4652 d^3 + 135656 d^4) .~\ee
The $d^3$ and $d^4$ terms only become important once $d \gae 0.02$, indicating that they are 
artifacts of the breakdown of our approximations. Therefore we see that the dominant behaviour of 
$\kappa(d)$ at small $d$ is $d^2$. Thus we can extrapolate 
$\kappa(d)$ in the form
\be{kdex} \kappa(d) \approx 1 + 110 d^2 .~\ee

\section{Cosmic Ray Catalysed vs. Spontaneous False Vacuum Decay}

	 In this section we give the condition for cosmic ray catalysed 
vacuum decay to be more important than spontaneous false vacuum decay. 
Cosmic rays can catalyse electroweak vacuum decay by scattering off 
 background protons and producing 
a virtual Higgs field of sufficient energy (which will be spherically symmetric 
in the centre of mass frame). As discussed earlier, the 
importance of the virtual scalar field is that it provides a source of energy for the 
growth of the true vacuum bubbles. For the virtual scalar field 
to be able to provide all of its energy to the critical bubble without introducing 
additional large suppression factors, it should be concentrated in
 a region at least as small as the radius of the critical bubble. 
In fact, for the thin-wall bubbles we are discussing here, the Compton wavelength of 
the virtual scalar field at the critical energy 
turns out to be generally less than or of the order of the width of the 
bubble wall, $ E_{c}^{-1} \lae w $. This follows because  
the condition $ E_{c}^{-1} \lae w $ is equivalent to
\be{compton} \frac{\epsilon}{V_{b}} \lae
\left( \frac{64 \sqrt{2} \pi}{3} \frac{e_{c}}{\lambda}\right)^{1/2}  ,~\ee
where $V_{b} \equiv \lambda v^{4}/4$ is the height of the potential barrier.
Thus with $e_{c} \approx 0.15$ this will be
satisfied if $ \epsilon \lae 3.7 V_{b}/\lambda$. Since $\epsilon$ is the 
splitting of the energy density of the vacuum states, which is less than or 
of the order of $V_{b}$ for the case of thin-wall bubbles,
we see that this will generally be satisfied in the thin-wall limit
we are considering.

		 The decay rate is largest for cosmic-ray collisions of centre 
of mass energy approximately equal to the critical bubble energy.
This is because, as discussed in the previous sections, the
 scalar field catalysed decay probability is highest for a scalar field of 
energy equal to the 
critical energy $ E_{c}$. Since the flux of cosmic rays rapidly 
decreases with increasing energy \cite{crf} whilst the 
probability of vacuum decay does not increase with energy 
once the scalar field energy is 
greater than the critical energy \cite{v1}, it follows that
the contribution to the vacuum decay rate 
from higher energy cosmic ray collisons will be negligible. 

		     The cross-section for a proton-proton collision to produce
 a Higgs boson of energy  $ E_{h}$ via gluon fusion is given by \cite{gf}
\be{xsect} 
\frac{d \sigma}{d E_{h}}(E_{cm}, E_{h}) =
 \frac{1}{\sqrt{E_{h}^{2}-m_{h}^{2}}} 
\frac{G_{F} \alpha_{s}^{2}}{288 \pi \sqrt{2}}\; 
x_{1} x_{2} G(x_{1}) G(x_{2})      ~,
\ee
 where $ G(x)$ is the gluon distribution function, $ x_{1} = 
(\frac{m_{h}}{E_{cm}}) e^{y}$, $ x_{2} = (\frac{m_{h}}{E_{cm}}) e^{-y}$ and 
$ e^{2 y} = (E_{h} + \sqrt{E_{h}^{2} - m_{h}^{2}})(E_{h}
- \sqrt{E_{h}^{2}-m_{h}^{2}})$. 
Thus if we are interested in producing a Higgs boson in an energy range 
$ \delta E_{h}$ around $ E_{h}$ for 
a given $ E_{cm}$, the cross-section is given by
\be{sigma} \nonumber
\sigma (E_{cm}, E_{h}) \approx \frac{ d \sigma}{d E_{h}}
 \delta E_{h}   ~,
\ee
which gives, for $ E_{h}^{2} \gg m_{h}^{2}$, 
\be{sigma2} 
\sigma (E_{cm}, E_{h}) \approx \frac{\delta E_{h}}{E_{h}} 
\frac{G_{F} \alpha_{s}^{2}}{288 \pi \sqrt{2}} \left(\frac{m_{h}}{E_{cm}}
\right)^{2} G(x_{1}) G(x_{2})      ~,
\ee
where $ x_{1} = \frac{2 E_{h}}{E_{cm}}$ and $ x_{2} =
 \frac{m_{h}^{2}}{2 E_{h} E_{cm}}$.
We will consider $ \delta E_{h} 
\approx E_{h}$ in the following.
 The centre of mass energy is $ E_{cm} = 
\sqrt{2 m_{p} E_{1}}$, where $ E_{1}$ is the cosmic ray energy 
and $ m_{p}$ the proton mass. Since it is best to produce a Higgs
 of energy $ E_{h}$ with the minimum cosmic ray energy $ E_{1}$,
we shall consider the cross-section for the case
 $ E_{cm} \approx E_{h}$, corresponding to cosmic rays of energy 
$ E_{1} \approx \frac{E_{h}^{2}}{2 m_{p}}$, . 
The total 
number of cosmic ray induced vacuum decay events during 
the history of the Universe is then given by   
\be{events} 
N(E_{h}) \approx  f(E_{1}) \sigma (E_{1}) P(E_{h}) 
 n_{p} t_{u} \delta E_{1} ~,
\ee
where $f(E)$ is the flux of cosmic ray protons of energy $E$, $ P(E_{h})$ 
is the probability of catalysed vacuum decay in 
a virtual scalar field of energy 
$ E_{h}$, $ n_{p} \approx 2\times 10^{78}$ is the number of protons in
 the Universe and $ t_{U} \approx 5\times 10^{41} \GeV^{-1}$ is the age of
 the Universe \cite{eu}. 
In this we have considered $ N(E_{h})$ to come mostly from a range 
$ \delta E_{1} \approx E_{1}$ around the minimum cosmic ray energy
 needed to
 produce a Higgs of energy $ E_{h}$; this is consistent with 
$\delta E_{h} \approx E_{h}$. 
The cosmic ray flux is given by (for $ E_{1} > 100~\GeV$) \cite{crf} 
\be{flux} 
f(E_{1}) \approx A \left(\frac{E}{100~ \GeV}\right)^{-\gamma} 
 \;\;\; ; \;\;\; \gamma = 2.75   ~,
\ee 
where $ A = 2.2\times 10^{-56}\GeV^{2}$. 
We define $ \chi(E_{h})$ by
$ N(E_{h})  = \chi (E_{h}) P(E_{h})$. From \eq{events} we find 
\be{chi} 
\chi(E_{h}) \approx e^{134 - \rho (E_{h},m_{h})}    ~,
\ee
where 
\be{define} 
e^{-\rho (E_{h},m_{h})} = 
G(2) G\left(\frac{m_{h}^{2}}{2 E_{h}^{2}}\right)
\left(\frac{m_{h}}{100 \GeV}\right)^{2} 
\left(\frac{100 \GeV}{E_{h}}\right)^{2.75}    ~.
\ee
Thus, with $ \eta$ defined by $ P(E_{h}) = e^{-\eta}$,
 we find that catalysed false vacuum decay will typically 
occur for $ \eta \lae 130$. 

	       Spontaneous false vacuum decay occurs at a rate 
per unit volume per unit time given by \cite{cea}
\be{gammas} 
\Gamma_{spont} \approx m_{h}^{4} \; \exp\left(-S_{4}\right)    ~,
\ee
where $ S_{4}\; (\equiv F(0,d))$ is the bounce action. 
Thus the probability of spontaneous 
vacuum decay during the history of the Universe is given by
\be{gammas2} 
\Gamma_{spont} t_{u} r_{u}^{3}  \approx 10^{175} e^{-S_{4}}~,
\ee
where $ r_{u} \approx t_{u}$ is the radius
 of the Universe and we have used $ m_{h} = 10^{2} \GeV$. Therefore
 spontaneous vacuum decay will occur if $ S_{4} \lae 403$. In terms of
 $ \kappa(d)$, since $ F(0,d)$  is equal to the bounce action 
$ S_{4}$ and since $ 
F(E_{c},d) \lae 130$ implies that cosmic ray 
catalysed false vacuum 
decay occurs, we see that catalysed false vacuum decay 
will be more important than spontaneous decay if 
\be{kappad} 
\kappa (d) \lae 2.0    ~.
\ee     
  
Using the extrapolation of $\kappa(d)$, \eq{kdex}, we find that 
$\kappa(d) \approx 2$ occurs at $d \approx 0.1$, well outside the thick-wall
limit where our results would become unreliable. (The initial radius of a bubble at the 
critical energy is $x_{c} = 0.67$ whilst the total thickness of the wall is 2$d$. Therefore the 
bubble wall width in this case is about 30$\%$ of its initial radius). 
Thus we can conclude that cosmic-ray catalysed false vacuum decay can 
only be important if the bubbles are not thick-walled. In terms of
the parameters of the potential \eq{pot}, the condition \eq{kappad} 
implies that $M^3\lae 0.1\lambda v^3$. 

	   Could false vacuum decay to thin-walled bubbles ever be cosmologically 
significant? The answer is almost certainly 
not, at least not for the case of a perturbative model.
 To see this, consider spontaneous vacuum decay. 
 For a real scalar field with a potential given by
\be{phipot} V(\phi) = \frac{m^2}{2} \phi^2 - \frac{\delta}{3} \phi^3
 + \frac{\lambda}{4} \phi^4   ,~\ee
to which most phenomenological models can be roughly fitted,
the bounce action is given by \cite{cea}
\be{baction} S_{4} = \frac{9}{4} \frac{m^2}{\delta^2} 
\tilde{S}_{4}(\tilde{\lambda})   ,~\ee
where $ \tilde{S}_{4}(\tilde{\lambda})$ is the rescaled bounce action as a 
function of $\tilde{\lambda}$ and $\tilde{\lambda} 
\approx \frac{9}{2} m^2\lambda/\delta^2$. 
Thin-wall bubbles correspond to $\tilde{\lambda} \gae 0.85$, for which $ d \lae 0.2$. In this case 
it may be shown that $ \tilde{S}_{4}(\tilde{\lambda}) \gae 2600$. Therefore 
\be{s4limit} S_{4} \approx \frac{\tilde{S}_{4}}{2 \lambda} \gae \frac{1300}{ \lambda}  ,~\ee
where we have used $ \delta^2 \approx \frac{9}{2} m^2 \lambda$ for the thin-wall limit.   
Since cosmologically significant spontaneous 
false vacuum decay requires that $S_{4} \lae 400$, we see that 
we would require $\lambda \gae 3.3$ in the thin-wall case. 
Since for the theory described by \eq{phipot} the 1-loop correction to the 
scalar self-interaction becomes larger than the tree-level contribution 
for $\lambda \gae 2$, 
 we see that thin-wall false vacuum decay could only be 
cosmologically significant if the scalar field were non-perturbative. 
Although this result is true for 
spontaneous false vacuum decay, the possible enhancement of the 
cosmic ray catalysed false vacuum decay rate 
over the spontaneous decay rate in the 
thin wall limit will not alter this conclusion, since the increase in 
$\tilde{S}_{4}(\tilde{\lambda})$ is far greater 
than the reduction in $\kappa(d)$ as we go to thinner bubbles.

\section{Conclusions}

	  In this paper we have attempted to address the question of whether false vacuum decay
catalysed by cosmic ray collisions could ever be an important constraint on models of
particle physics. We have used the Landau-WKB method, including the effects of 
finite bubble wall width, to calculate the rate of cosmic ray
catalysed false vacuum decay relative to that for spontaneous decay. 
We find that cosmic ray catalysed decay is the dominant process for sufficiently 
thin-walled bubbles. However, for the case of perturbative 
models with cosmologically significant false vacuum decay
rates, which almost certainly decay via thick-walled bubbles, our results strongly suggest that 
spontaneous false vacuum decay via tunnelling will always be the dominant decay mode.

We should also note that our results indicate that vacuum decay is very
unlikely to be catalysed by particle collisions in accelerators; the total
luminosities involved are simply far too low. In the very early universe the 
situation might however be different. It is conceivable that, for
a certain range of the potential parameters, catalysed
decay could be more important than thermal fluctuation over the barrier or
the zero temperature tunnelling. Such a possibility could constrain particle physics
models in an interesting way and merits further study.

Our results are based on extrapolating the 
cosmic ray catalysed decay rate from the small values of wall thickness 
for which our calculation is valid to larger thicknesses
(although still well outside the thick-wall limit). This extrapolation 
indicates that catalysed 
false vacuum decay becomes less important than spontaneous decay once the 
bubbles produced by cosmic ray collisions have a thickness greater than 
about 30$\%$ of their
initial radius. It is possible that the behaviour of the catalysed decay rate as a function of wall 
thickness could begin to deviate from the thin-wall behaviour at such thicknesses, although we do not
expect our conclusions to change significantly. However, it would be valuable to have a more 
accurate calculation for wall thickness which are between the thin and thick wall regimes. 
 To go beyond the 
thin wall approximation using the Landau-WKB method would require an action describing 
the dynamics of thick(er)-walled bubbles as a function of a single variable 
$R$. This could 
be done, for example, by considering numerical solutions of the $ \phi$ equations 
of motion and then deducing an Ansatz for $ \phi$ which is a
 reasonable approximation to the observed behaviour. Substituting 
this into the scalar field action would, in principle, allow for the derivation of a thick wall 
action as a function of R, to which the Landau-WKB method could be applied. 

	  The analytical Landau-WKB approach we have considered here 
is complementary to the numerical approach as
followed by Kuznetsov and Tinyakov \cite{tea}. The trend of our results with increasing wall width 
agrees with that suggested by their thick wall bubble results.  
We believe that the Landau-WKB approach 
 can give us additional insights into the physics of catalysed false-vacuum decay
and provides a powerful approach to the problem.

\section*{Acknowledgements}
K.E. wishes to thank Kimmo Kainulainen for a useful discussion on branch
cuts. This work has been supported by the Academy of Finland and 
the EU contract ERBFMBICT960567. 
\newpage
\section*{Appendix. Landau-WKB Method in Quantum Mechanics}

In his original discussion, Landau \cite{ll} was concerned 
with calculating the matrix element of an operator $f(x)$ 
between two states, $ \psi_{1}$ 
and $ \psi_{2}$, corresponding to particles of energies $ E_{1}$ and
 $ E_{2}$ ($ E_{2} > E_{1}$) reflecting from a potential barrier
 $U(x)$. The presence of the potential barrier makes it possible for 
 the matrix element between the two energy eigenstates to be non-zero. 
 The standard WKB wavefunctions for this system are given by 
(with $ \hbar =1$)
 \bea{psi1} 
\psi_{1} &=& \frac{C_{1}}{2 \sqrt{|p_{1}|}} 
\exp\left[-\left| \int_{a_{1}}^{x} p_{1} 
\; dx\right| \right]  \, \, \, \, \, x < a_{1} \nn
\psi_{1} &=& \frac{C_{1}}{\sqrt{p_{1}}} \;
\cos \left(\int_{a_{1}}^{x} p_{1} \;dx - \frac{\pi}{4} 
\right)  \, \, \, \, \, x > a_{1}  \nn
\eea
and $ \psi_{2} = \psi_{2}^{+} + \psi_{2}^{-}$ where 
\bea{psi2} 
\psi_{2}^{+} &=& \frac{-i C_{2}}{2 \sqrt{|p_{2}|}}\; 
\exp\left[\left| \int_{a_{2}}^{x} p_{2} \; dx 
\right| \right]  \, \, \, \, \, x < a_{2}  ~\nn
\psi_{2}^{+} &=& \frac{C_{2}}{2 \sqrt{|p_{2}|}} \;
\exp \left[\int_{a_{2}}^{x} p_{2} dx - 
\frac{i \pi}{4} \right]  \, \, \, \, \, x > a_{2}  ~\nn
\eea
and $ \psi_{2}^{-} = (\psi_{2}^{+})^{*}$. 
In this the full $ \psi_{2}$ wavefunction has been set to zero
 at $ x < a_{2} < a_{1}$, where it does not contribute significantly to 
the matrix element (both $ \psi_{1}$ and 
$ \psi_{2}$ being exponentially suppressed in this region). The
 matrix element of an operator $f$ between $ \psi_{1}$ and $ \psi_{2}$ is then given by
$f_{12} = f_{12}^{+} + f_{12}^{-}$, where
\be{kaava1}
f_{12}^{+} = \int_{-\infty}^{\infty} \psi_{1} f \psi_{2}^{+} \; dx   ~
\ee
and $ f_{12}^{-} = (f_{12}^{+})^{*}$. 
The WKB wavefunctions are not valid close to the 
turning points $ a_{1}$ and $ a_{2}$. However, by continuing
$x$ to the upper half of the complex plane
 these points may be avoided. The wavefunctions in the upper half-plane may then 
be written as
\bea{psiit}
 \psi_{1} &=& \frac{C_{1}}{2 [i p_{1}]^{1/4}}
\exp\left[ i\int_{a_{1}}^{x} p_{1} dx \right]    \nn
\psi_{2}^{+} &=& \frac{-i C_{2}}{2  [i p_{2}]^{1/4}}
 \exp \left[-i\int_{a_{2}}^{x} p_{2} dx \right]~.
\eea 
The matrix element is then given by 
\bea{matrelem}
 f^{+}_{12} &=& \frac{-i C_{1} C_{2}}{4 }
\int \frac{f(x) dx}{[-p_{1}p_{2}]^{1/4}}   \nn
&\times& 
\exp\left[ i\int_{a_{1}}^{x}  p_{1}dx^{'}   -
 i\int_{a_{2}}^{x} p_{2}dx^{'} \right]  
    ~. 
\eea           
By the residue theorem, the matrix element will be determined by the points
 in the complex plane ("transition points") 
where the true wavefunctions are singular, which 
corresponds to the points at which the exponential factor is
 stationary as a function of $x$, which we denote by $ x^{*}$. 
This follows because at these points the canonical momenta for two
 different energies are equal, which implies that the momenta are independenent  
of the energies of the states. Therefore the potential and 
so the wavefunctions at these points must be singular. 
The integral, up to a relatively unimportant prefactor, 
will then simply equal the 
exponential factor evaluated at $ x^{*}$. Thus 
\be{expfact} 
f_{12}^{+} \sim \exp \left[ \int_{a_{1}}^{x^{*}} \sqrt{2 m (U-E_{1})}\; 
dx   -
\int_{a_{2}}^{x^{*}} \sqrt{2 m (U-E_{2})}\; dx \right]   ~.
\ee
where we have used $ p_{i} = -i \sqrt{2 m (U-E_{i})}$ and 
where the square roots are taken to be positive on the real axis for
 $ x < a_{2}$. 
Since the complex part of the exponent contributes only an 
insignificant phase factor, we can write 
$ f_{12} = f_{12}^{+} + f_{12}^{-}$ as 
\be{viimonen} 
f_{12} \sim \exp\left[~\Re\left(\int_{a_{1}}^{x^{*}} 
\sqrt{2 m (U-E_{1})}\; dx
 - \int_{a_{2}}^{x^{*}} \sqrt{2 m (U-E_{2})}\; dx \right) \right]  ~.
\ee
This expression is true so long as the states in the matrix element
 are "strongly different",
 meaning that the exponential suppression factor is large.

\end{document}